\documentclass[10pt]{article}
\usepackage{amsmath}
\usepackage{amsfonts}
\usepackage{amssymb}

\newcommand{\smaq}{\left[ \begin{smallmatrix}}
\newcommand{\smat}{\left( \begin{smallmatrix}}
\newcommand{\smcq}{\end{smallmatrix}\right]}
\newcommand{\smct}{\end{smallmatrix}\right)}
\newcommand{\transp}[1]{{}^t\!#1}

\DeclareMathOperator{\modular}{Sp}

\DeclareMathOperator{\Imm}{Im}

\newcommand{\ZZ}{{\bf Z}}

\newcommand{\RR}{{\bf R}}
\newcommand{\CC}{{\bf C}}

\newcommand{\HH}{{\bf H}}

\newcommand{\be}{\begin{equation}}
\newcommand{\ee}{\end{equation}}
\newcommand{\bes}{\begin{equation*}}
\newcommand{\ees}{\end{equation*}}
\newcommand{\eqn}{\begin{eqnarray}}
\newcommand{\feqn}{\end{eqnarray}}
\newcommand{\arr}{\begin{eqnarray*}}
\newcommand{\farr}{\end{eqnarray*}}

\title{Construction of chiral superstring measure$^\dagger$}

\author{F. Dalla Piazza}

\begin{document}
\maketitle

{\small {\it Dipartimento di Matematica e Fisica, Universit\`a dell'Insubria, 22100
Como, Italy, and I.N.F.N. sezione di Milano, via Celoria 16, 20100 Milano, Italy.}}

\begin{abstract}
The mathematically rigorous definition and construction of the amplitudes in superstring theory is still an open problem. Here, we describe some recent development in the construction of the superstring measures in $g=3,4$ and we point out some aspects that are not yet clear.
\end{abstract}

{\small \it
\noindent $^\dagger$ This paper is substantially a review of the papers \cite{CP,CDG,CDG2,DG}, based on a talk given by the author at
``XVIII Congresso SIGRAV, General Relativity and Gravitational Physics, Cosenza, 22-25 Settembre 2008''.}

\section{Introduction}
The mathematically rigorous definition and construction of scattering amplitudes in superstring theory is still an open problem. The same topic in the ambit of bosonic strings was solved in the eighties and the solution finds a strong geometric foundation in the Mumford's theorem. The problem of defining the amplitudes is of central interest in superstring theory in the perturbative approach because, as for any other QFT, through the perturbative expansion one constructs the theory itself. Indeed, unless using an assiomatic approach, we are able to define consistently a quantum field theory just by means of the perturbative approach.
In this paper we will describe some our results on the construction of the chiral measure in superstring theory for $g>2$. The proposal for a candidate of the superstring measures at genus three and four and the proof, under certain assumption, of the uniqueness of these measures are our major results. As we will see, the starting point is a result of D'Hoker and Phong, which permitted us to propose a reasonable guess for the proprieties the superstring measures should satisfy. This ``conjectural" approach allows us to find an explicit expression for the amplitudes at genus three and four.
Our construction makes use of the theory of modular forms and of the theory of representations of finite groups. We will see that the amplitudes are particular modular forms with respect to a certain finite subgroup of the symplectic group and we will be able to express them in terms of theta functions.
Before describing the construction of the superstring measures  we will recall the notion of the amplitudes in superstring theory, their definition in the path integral formalism and the role played by the complex geometry of the Riemann surfaces.
\\

\section{Amplitudes in string theory}
In this section we introduce some aspects of string theory in the perturbative approach and define string amplitudes in the path integral formalism. We first overview the more rigorous results for the bosonic string, and then generalize this construction to the supersymmetric case. In both cases we will present the construction of the vacuum-to-vacuum amplitudes in the NRS formalism, and consider closed strings only.

\subsection{The bosonic case} \label{bosonic}
The bosonic string theory in Minkowski space time is defined by the Polyakov action on a Riemann surface $\Sigma_g$ of genus $g$:
\be \label{actionbos}
I(X,h)=\frac {1}{4\pi \alpha'} \int_{\Sigma_g} d^2z \sqrt h \ h^{ab} \partial_a X \cdot \partial_b X,
\ee
where the $X$'s are immersion functions from the Riemann surface $\Sigma_g$ to the traget space $\RR^d$, $h$ are the metrics on the Riemann surface and $z$'s are the coordinates on $\Sigma_g$.
Then, one defines the partition function as:
\be
Z_g^{bos}=\int [Dh_{ab}] [DX] \exp (-I(X,h)),\label{partition}
\ee
the notation $[\cdots]$ stand for the functional measure, i.e. the integration is formally performed over all the metrics on $\Sigma_g$ and over all immersions $X$. The complete partition function involves the sum over all genera of the Riemann surfaces.
Employing the huge symmetry of the classical action one can reduce the path integral to a finite dimensional integral over the moduli space $\mathcal{M}_g$ of the Riemann surface $\Sigma_g$. Actually, the action \eqref{actionbos} has as symmetry group the semidirect product $G=\mbox{Weyl}(\Sigma)\ltimes\mbox{Diff}(\Sigma)$ of the Weyl transformations and the group of diffeomorphisms of the Riemann surface. Thus, the moduli space for conformal class of Riemann surfaces is defined as $\mathcal{M}_g=M/G$, where $M$ is the set of all possible Riemannian metrics over $\Sigma_g$. The moduli space turn out to be a finite complex manifold. The integration over the $X$'s fields can be easly performed as a Gaussian integral and can be computed in terms of the determinant of the Laplacian associated to the metric $h_{ab}$. For details see, for example, \cite{nov}. In the computation of the determinant one has to pay attention to the presence of the zero modes that must be dropped out. This breaks the conformal invariance and the procedure becomes anomalous unless the dimension of the space time is $D=26$. In this way one exploits the spectral proprieties of the Laplacian, but one can also follows a more geometrical approach in which the complex geometry underlying the Riemann surface is taken into account. This last procedure has been followed by Belavin and Knizhnik, Belison and Manin and also by D'Hoker and Phong and allows to express the path integral formula for the partition function  in terms of global geometric objects. The starting point is a theorem due to Mumford which asserts that the linear bundle $U=K\otimes\lambda^{-13} $ is a holomorphically trivial bundle over $\mathcal{M}_g$, where $K$ is the canonical bundle over $\mathcal{M}_g$, i.e. the highest wedge power of the cotangent bundle, and $\lambda$ is the Hodge bundle over $\mathcal{M}_g$, i.e. the highest wedge power of the holomorphic cotangent bundle. As a consequence $U$ admits an essential unique holomorphic global section $\psi_g$, the Mumford section. The section $\psi_g$ is nonvanishing everywhere and meromorphic at infinity with a pole of order two. The Belavin-Knizhkin theorem implies that the bosonic partition function can be written in term of the square modulus of the Mumford section. Moreover Manin has been able to write down the partition function in terms of theta functions. This expression for the amplitude measures in terms of global objects seems to be the more useful for our construction of the superstring measure at $g=3,4$ and for a possible generalization to higher genus. Moreover, this geometric approach provides a rigorous derivation for the expression of the partition function. The only undesirable aspect is the divergence due to the pole at infinity of the Mumford section and this can be imputed to the presence of the tachyon in the bosonic spectrum. This difficulty can be overcomed in superstring theory by means of the GSO projection.

\subsection{The supersymmetric case} \label{super}
The bosonic string can be generalized to the supersymmetric case. On a Riemann surfaces $\Sigma_g$ of genus $g$ there are $2^{2g}$ different spin structures and for any choice of them we can define a spinor field over the surface. For any fixed choice of the spin structure\footnote{In what follows we will use the capital $\Delta$ for spin structures in any genus $g$ and a small $\delta$, in analogy with D'Hoker and Phong, for the special case $g=2$.} $\Delta$ the superstring action is given by \cite{nov}:
\begin{align} \label{superaction}
I_{\Delta}&=\frac 1{4\pi \alpha'} \int_\Sigma d^2z \sqrt {h} [\frac 12 h^{\alpha\beta} \partial_{\alpha} x^\mu \partial_{\beta} x_\mu
-\frac i2\psi^\mu \gamma^\alpha D_\alpha \psi_\mu -\frac 12\psi^\mu \gamma^a \gamma^\alpha \chi_a \partial_\alpha x_\mu \\  \nonumber
&+\frac 18 \psi^\mu \gamma^a \gamma^b \chi_a (\chi_b \psi_\mu)]+\lambda {\mathcal{X}}(\Sigma),
\end{align}
where $\chi_a^\alpha$ is the gravitino, the superpartner of the metric $h_{\alpha\beta}$, $\psi^\mu$ are Majorana spinors, superparteners of the coordinate $x^\mu$, and $\mathcal{X}$ is the Euler characteristic.
Beyond the symmetries given by diffeomorphism and Weyl transformations the action \eqref{superaction} is also left invariant by supersymmetric and super-Weyl transformations. As for the bosonic case, for each fixed spin structure $\delta$, one defines the partition function:
\be \label{superpartit}
Z_\Delta=\int [Dh_{\alpha\beta}] [D\chi_a] [Dx^\mu] [D\psi^\mu] \exp (-I_\Delta).
\ee
Also in supersymmetric case the cancellation of anomaly fixes the space time dimensions and in this case one obtains that the theory is consistent only if a ten dimensional space time is taken into account.
Performing computations in the path integral approach, one obtains for the partition function an expression that, although conformally invariant, contains some ambiguities. Actually this expression is not completely independent by the choice of the parametrization of the moduli. Any change in the parametrization adds to the integral some boundary terms which should vanish, but which really do not. We will not investigate further the issues connected to this ambiguity, a detailed analysis can be found in \cite{mooremor,atrasen,atmoosen}.
D'Hoker and Phong in a series of remarkable papers \cite{DP1,DP2,DP3,DP4} have made a proposal to solve the ambiguity. In superstring theory, actually, one deals with super Riemann surfaces and their moduli space is a superspace. D'Hoker and Phong suggested that the ambiguity should be imputed to a wrong choice of the parametrization of the bosonic part of such super moduli space. Indeed, usaually, one choses the metric as the bosonic part and the gravitino as the fermionic one. Suppose that a particular slice is selected by the choice of the metric and the gravitino $(h_{\alpha\beta},\chi_\alpha)$. A key point in the computation of the superstring amplitudes is the apparently natural projection
\be \label{proj}
(h_{\alpha\beta},\chi_\alpha)\rightarrow h_{\alpha\beta}
\ee
that allows to get rid, after an integration, of the fermionic degree of freedom. However, under a supersymmetric transformation, one obtains a new slide $(\tilde{h}_{\alpha\beta},\tilde{\chi}_\alpha)$. If the metric should be a good choice for the bosonic components then the natural projection \eqref{proj} would be supersymmetric preserving, i.e. the supersymmetric transformation should induce a diffeomorphism between the two metric $h_{\alpha\beta}$ and $\tilde{h}_{\alpha\beta}$. But this, in general, do not happens: $h_{\alpha\beta}$ and $\tilde{h}_{\alpha\beta}$ are not related by a bosonic symmetry:
\bes
\begin{array}{ccc}
(h_{\alpha\beta},\chi_\alpha) &\sim & (\tilde{h}_{\alpha\beta},\tilde{\chi}_\alpha) \\
\downarrow   &  &\downarrow \\
h_{\alpha\beta} & \nsim & \tilde{h}_{\alpha\beta}.
\end{array}
\ees
The main idea of D'Hoker and Phong is to substitute the metric with the period matrix associated to the Riemann surfaces considered. In this way they obtain a slice parametrization given by the notion of the super period matrix that is supersymmetric and has not the same problems as the metric. In a long series of papers they has been able to perform the computations just for the genus two case, but they argued that their argument is general and it should hold for any genus $g$. Finally, they found, for $g=2$, a nice and well defined expression for the amplitude as a polynomial in the theta constants:
\begin{equation}
Z_2=\int_{{\mathcal{M}}_2} (\det Im \Omega)^{-5} \sum_{\delta \delta'} c_{\delta \delta'} d\mu [\delta] (\Omega) \wedge \overline {d\mu [\delta'] (\Omega)}, \label{1punto1}
\end{equation}
where $\Omega$ is the period matrix, $\delta$ is an even spin structure\footnote{At genus two there are ten different even spin structures and each one can be written in two different ways as sum of three different odd spin structures $\nu$.}, $c_{\delta \delta'}$ are phases realizing the right GSO projection and
\begin{eqnarray}
&& d\mu [\delta] (\Omega)=\frac {\theta[\delta](0,\Omega)^4 \Xi_6 [\delta](\Omega)}{16\pi^6 \psi_{10}(\Omega)}\prod_{I\leq J} d\Omega_{IJ},\\
&& \Xi_6[\delta](\Omega):=\sum_{1\leq i<j\leq3}\langle\nu_i|\nu_j\rangle\prod_{k=4,5,6}\theta[\nu_i+\nu_j+\nu_k](\Omega,0)^4\ ,
\end{eqnarray}
where each even spin structure is written as a sum of three distinct odd spin structures $\delta=\nu_1+\nu_2+\nu_3$, whereas and $\nu_4$, $\nu_5$, $\nu_6$
denote the remaining three distinct odd spin structures,
\begin{eqnarray*}
\langle\kappa|\lambda\rangle:=e^{\pi i(a_\kappa\cdot b_\lambda-b_\kappa\cdot a_\lambda)},\qquad\ \kappa=
[{}^{a_\kappa}_{b_\kappa}], \quad \lambda=[{}^{a_\lambda}_{b_\lambda}]
\end{eqnarray*}
is a sign and $\theta[\delta](\Omega,0)$ are the theta constants, see section \ref{theta}. For example: 
\begin{align*}
\Xi_6\smaq 1 & 0 \\ 0 & 0 \smcq&=\theta\smaq 0 & 0 \\ 0 & 1 \smcq^4 \theta\smaq 0 & 0 \\ 1 & 0 \smcq ^4\theta \smaq 0 & 1 \\ 0 & 0 \smcq^4 -\\
& - \theta\smaq 0 & 0 \\ 0 & 0 \smcq^4 \theta\smaq 0 & 0 \\ 1 & 1 \smcq^4 \theta \smaq 0 & 1 \\ 1 & 0 \smcq^4 + \theta\smaq 1 & 0 \\ 0 & 1 \smcq^4 \theta\smaq 1 & 1 \\ 0 & 0 \smcq^4 \theta \smaq 1 & 1 \\ 1 & 1 \smcq^4.
\end{align*}
Finally, $\psi_{10}=\prod_{\delta\ {\rm even}} \theta[\delta](0,\Omega)$ is the Igusa form.
D'Hoker and Phong claimed that the expression \eqref{1punto1}, rigorously proved for $g=2$, should indeed be true for any genus $g$:
\be
Z_g=\int_{{\mathcal{M}}_g} (\det Im \Omega)^{-5} \sum_{\Delta \Delta'} c_{\Delta \Delta'} d\mu [\Delta] (\Omega) \wedge \overline {d\mu [\Delta'] (\Omega)},
\ee
where now $\Delta=[{}_a^b]$, $a,b\in\ZZ_2^g$, are the even spin structures (or theta characteristics) at genus $g$. Moreover, the measure $d\mu[\Delta](\Omega)$ for the string amplitudes at genus $g$ is expected to factorize as the bosonic measure times a suitable form: $d\mu[\Delta] (\Omega) =d\mu_{BOS} (\Omega) \Xi_8 [\Delta]$.
As explained in section \ref{bosonic}, the bosonic measure is a well defined object, nevertheless  it can be written in term of theta functions only up to genus four. The forms $\Xi_8[\Delta]$ have to be determined. D'Hoker and Phong in \cite{DP5,DP6} suggested precise ans\"atze for the measure at genus 3. The genus
3 bosonic measure is given by
$$
{\rm d}\mu^{(3)}_B\,=\,\frac{c_3}{\Psi_{9}(\Omega)} \prod_{I\leq J}\,{\rm d}\Omega_{IJ},
$$
where $\Psi_{9}^2(\Omega)$ is a Siegel modular form of weight $18$ for $Sp(6,\ZZ)$.
By analogy with the $g=2$ case, they proposed that the genus three chiral superstring measure should be of the form
$$
{\rm d}\mu[\Delta]\,=\, \frac{\theta[\Delta](0,\Omega)^4\Xi_6[\Delta](\Omega)}
{8\pi^4\Psi_{9}(\Omega)}\prod_{I\leq J}\,{d}\Omega_{IJ}, \qquad \Xi_8[\Delta]\equiv \theta[\Delta](0,\Omega)^4\Xi_6[\Delta](\Omega)
$$
and they gave three constraints on the functions
$\Xi_6[\Delta](\Omega)$. Actually, the forms $\Xi_6[\Delta]$ turn out to be particular modular forms belonging to a huge vector space. D'Hoker and Phong were not able to find functions satisfying these constraints. This failure must be imputed not just to the lack of a systematic procedure to study this space, but mainly to the fact that such forms do not exist, as we proved \cite{CDG} using the theory of representation of finite groups. We used analogous techniques in \cite{CP} to reinterpret the result of D'Hoker and Phong at genus two.

Our starting point in the construction of the supersting amplitudes is the assumption of the validity of \eqref{1punto1} at any genus, although its proof is still an open problem. 
We recall here that the bosonic measure is strongly supported by global issues in algebraic and complex geometry that essentially lie in Mumford theorem, instead for superstring theory one deals with the much less supported expression \eqref{1punto1}.
Morozov discussed two different approaches to solve the problem of superstring measures. The first is by direct integration of odd moduli after holomorphic factorization, as done by D'Hoker and Phong for the genus two case. The second one is to start from some reasonable guesses  for the measures, based on general consideration, and then use these ans\"atze to determine their explicit form. We will follow this second approach. In order to expose our construction, we now present the necessary mathematical instruments.

\section{Symplectic group and modular forms}
The symmetries of the action of the superstring theory are reflected by precise modular proprieties of the superstring measures. These proprieties are related to the way the forms $\Xi_8[\Delta]$ transform under the action of the symplectic group.

The symplectic group $\modular(2g,\ZZ)$ is the group of $2g\times 2g$ matrices that fix the symplectic form $E=\smat 0 & I \\ -I & 0 \smct$, $ME\,{}^t\!M=E$
for all $M\in \modular(2g,\ZZ)$.
We will focus on certain finite soubgroups $\Gamma\subseteq\modular(2g,\ZZ)$.

The Siegel upper half space, $\HH_g$, is the space of complex $g\times g$ symmetric matrices with positive imaginary part. We can see $\HH_g$ as a higher dimension generalization of the half upper complex plane:
\be
\HH_g:=\{\Omega\in M_g(\CC):\; {}^t\Omega=\Omega,\;\Imm(\Omega)>0\}.
\ee
The action of the group $\modular(2g,\ZZ)$ on $\HH_g$ is:
\bes
M\cdot \Omega\,:=\,(A\Omega+B)(C\Omega+D)^{-1},
\ees
for $M=\smat A & B \\ C & D \smct \in \modular(2g,\ZZ)$. The period matrix of a Riemann surface belongs to the Siegal upper half plane, but in general not all the points of $\HH_g$ are the period matrix of a Riemann surfaces (this is strictly true for $g\leq 3$). Moreover, Torelli's theorem asserts that a Riemann surfaces $\Sigma_g$ is completely determined by its period matrix.
 
A Siegel modular form $f$ of genus $g$ and weight $k$ with respect to the group $\Gamma\subseteq\modular(2g,\ZZ)$ is a function on the Siegel upper half space of genus $g$ such that:
\begin{itemize}
\item
$f$ is a holomorphic function on $\HH_g,\quad f:\HH_g\rightarrow\CC$
\item
$f$ transforms as $f(M\cdot \Omega)\,=\,\det(C\Omega+D)^kf(\Omega)
\qquad\forall M\in\Gamma,\quad\Omega\in\HH_g$,
\end{itemize}
plus, for $g=1$, the requirement that $f$ is holomorphic at infinity.

\subsection{Theta constants and characteristic} \label{theta}
For $\Omega\in\HH_g$ and $z\in\CC^g$, (classical) theta functions are defined by the series:
\be
\theta[\Delta](\Omega,z)\,:=\,\sum_{m\in\ZZ^g}
\,e^{\pi i({}^t(m+a/2)\Omega(m+a/2)+2{}^t(m+a/2)(z+b/2)},
\ee
with $[\Delta]=[{}^a_b]$, $a=(a_1,\ldots,a_g),\;b=(b_1,\ldots,b_g)$, $a_i,b_i\in\{0,1\}$. The array $[\Delta]$ is called theta characteristic (or spin structure). A characteristic $\Delta$ is even or odd if $\sum_{i=1}^ga_ib_i$ is equal to 0 or 1 $\pmod 2$ respectively. For genus $g$ there are $2^{2g}$ different characteristics, $2^{g-1}(2^g+1)$ are even and $2^{g-1}(2^g-1)$ are odd. It can be proved that theta functions are even or odd in $z$ if their characteristic is even or odd.  When evaluated in $z=0$ these special functions are called theta constants. Thus, theta constants with odd characteristic vanish. There is a natural (affine) action of the symplectic group on the theta characteristics, see \cite{DG} for details, given by:
\be \label{transform}
\begin{pmatrix}A&B\\C&D\end{pmatrix}\cdot [{}^a_b]\,=\,
\left(\begin{array}{cc} D&-C\\-B&A\end{array}\right)
\left(\begin{array}{c}\transp{a}\\ \transp{b}\end{array}\right)
\,+\,
\left(\begin{array}{c}(C\transp{D})_0\\(A\transp{B})_0\end{array}\right)
\quad \mbox{mod}\;2
\ee
where, for a $g\times g$ matrix $N$,  $N_0=(N_{11},\ldots,N_{gg})$ is the diagonal of the matrix $N$. Then one has (\cite{Igusa}, V.1, Corollary) that:
\be
\theta[M\cdot\Delta](M\cdot\Omega)\,=\,
\kappa(M)e^{2\pi i\phi_{\Delta}(M)}\det(C\Omega+D)^{1/2}\theta[\Delta](\Omega),
\ee
for all $M\in Sp(2g,\ZZ)$ and with $\kappa(M)e^{2\pi i\phi_{\Delta}(M)}$ an eight-root of unity. We call \eqref{transform} the transformation formula. Using this formula we can prove that theta constant are ``almost modular''\footnote{For the presence of the constant $\kappa(M)$ we used the expression ``almost modular''. An expression for $\kappa(M)$ is available \cite{Igusa} in the case of squared theta constants $\theta[\Delta](\Omega)^2$.} forms of weight $1/2$ for a suitable finite subgroup of $\modular(2g,\ZZ)$.
Theta constants are a powerful tool to build up modular forms. It is convenient to define the $2^g$ (second order) theta constants:
\be
\Theta[\sigma](\Omega)\,:=\, \theta[{}^\sigma_{0}](2\Omega,0),\qquad
[\sigma]=[\sigma_1\;\sigma_2\;\ldots\;\sigma_g],\;\sigma_i\in\{0,1\},\;
\Omega\in\HH_g.
\ee
The classical theta constants and the second order theta constants span the same vector space, but the advantage of using the second order theta constants is that for $g=1,2$ they are all independent (i.e. there are no algebraic relations among them) and for $g=3$ there is just one relation among the eight second order theta constants. This relation is given by the locus of the zeros of a degree sixteen polynomial in eight variables, $F_{16}(\cdots\Theta[\sigma]\cdots)=0$. 
It happens that for $g\leq 3$ any modular form of weight $2k$ can be written as a homogeneous polynomial of degree $4k$ in the (second order) theta constants. Moreover, for $g<3$ this polynomial is unique and for $g=3$ it is unique if its degree is less then 15 otherwise it is
unique up to the addition of $F_{16}G_{4k-16}$, where $G_{4k-16}$
is any homogeneous polynomial of degree $4k-16$ in the theta constants and $F_{16}$ is the homogeneous polynomial (the unique one!)  of degree 16 in (second order) theta constants identically vanishing. For $g>3$, modular forms which can not be expressed as polynomial in theta constant can exist.
These considerations are crucial to prove the uniqueness of the superstring measures for $g=1,2,3$ and, in a weakened form, for $g=4$.

\section{The measures}
In section \ref{bosonic} and \ref{super} we reviewed the construction of the bosonic string and superstring measures respectively and we anticipated that they are intimately connected. In Table 1 we report the well known expressions for the lower genus measures $d\mu$. They was first computed in \cite{BKMP,M,Mor}.
\begin{table}[!h] \label{tab:measures}
\begin{center}
\bes
\begin{array}{ccc}
\hline
g & \mbox{Bosonic measure} & \mbox{Superstring chiral measure} \\
\hline
1 & {\rm d}\mu_B\,=\,
\frac{1}{(2\pi)^{12}\eta^{24}(\tau)}
{\rm d}\tau & {\rm d}\mu[\Delta]\,=\,
\frac{\theta[\Delta]^4(\tau)}{2^5\pi^4\eta^{12}(\tau)}
{\rm d}\tau \\[0.3em]
2 & {\rm d}\mu_B\,=\,\frac{c_2}{\Psi_{10}(\tau)}\prod_{i\leq j}\,{\rm d}\tau_{ij}
 & {\rm d}\mu[\Delta]\,=\,
\frac{\theta[\Delta]^4(\tau)\Xi_6[\Delta](\tau)}
{16\pi^6\Psi_{10}(\tau)}\prod_{i\leq j}\,{\rm d}\tau_{ij} \\[0.3em]
3 & {\rm d}\mu_B\,=\,\frac{c_3}{\Psi_{9}(\tau)}
\prod_{i\leq j}\,{\rm d}\tau_{ij}
 & ? \\
4 & {\rm d}\mu_B(g=4) & ? \\
\hline
\end{array}
\ees
\caption{Bosonic and superstring measures. The two question marks show the object of our construction.}
\end{center}
\end{table}
Our analysis focus on the construction of the genus three and four superstring measures, which are indicated in Table 1 by two question marks. The function $\eta$ is the Dedekind function and can be expressed in terms of the genus one theta constants as $\eta^{12}=\theta[{}_0^0]^4\theta[{}_0^1]^4\theta[{}_1^0]^4$, $c_2$ and $c_3$ are suitable constants. Although $d\mu_{B}$ for $g=4$ is known explicitly we do not report it here for brevity, see for example \cite{Mor}. It is clear that the superstring measures for $g=1,2$ can be expressed by the bosonic measure times a suitable form:
\begin{align*}
{\rm d}\mu^{(1)}[\Delta^{(1)}]\,&=\,c_1'
\theta[\Delta^{(1)}](\Omega^{(1)})^4\eta(\Omega^{(1)})^{12}\,{\rm d}\mu_B^{(1)}  \\
{\rm d}\mu^{(2)}[\Delta^{(2)}]\,&=\,c_2'
\theta[\Delta^{(2)}](\Omega^{(2)})^4\Xi_6[\Delta^{(2)}](\Omega^{(2)})\,
{\rm d}\mu_B^{(2)},
\end{align*}
in the apices we indicate the genus, when it is not clear from the contest and we will omit them every time it is possible. To lighten the notation we will often omit the argument $\Omega$. This factorization suggests a general form for the superstring measures:
\be
{\rm d}\mu^{(g)}[\Delta^{(g)}]\,=\,
c_g\Xi_8^{(g)}[\Delta^{(g)}](\Omega^{(g)}){\rm d}\mu^{(g)}_B,
\ee
where the forms $\Xi_8^{(g)}[\Delta^{(g)}](\Omega^{(g)})$ are, as we will see, suitable modular forms of weight 8 with respect to a certain finite subgroup of the symplectic group.


\section{Ans\"atze for the forms $\Xi_8^{(g)}[\Delta^{(g)}]$}
As promised in section \ref{super}, we give here some reasonable ans\"atze for the forms $\Xi_8^{(g)}[\Delta^{(g)}]$ which will provide unique solutions. We impose three constraints which are substantially the same of those of D'Hoker and Phong, the only difference being in the second one (the transformation constraint):
\begin{enumerate}
\item
the functions $\Xi_8[\Delta]$ are holomorphic on $\HH_g$;
\item
transformation condition under the action of $\modular(2g,\ZZ)$:
\bes
\Xi_8[M\cdot\Delta](M\cdot\Omega)\,=\,
\det(C\Omega+D)^8\Xi_8[\Delta](\Omega),
\ees
for all $M\in\modular(2g,\ZZ)$;
\item
restriction condition on 'reducible' period matrices:
\bes
\Xi_8[{}^{a_1\ldots a_k\,a_{k+1}\ldots a_g}_{b_1\ldots b_k\,b_{k+1}\ldots b_g}](\Omega_{k,g-k})\,=\,
\Xi_8[{}^{a_1\ldots a_k}_{b_1\ldots b_k}](\Omega_k)
\Xi_8[{}^{a_{k+1}\ldots a_g}_{b_{k+1}\ldots b_g}](\Omega_{g-k}),
\ees
where:
\bes\Delta_{k,g-k}\,:=\,\left\{\Omega_{k,g-k}\,:=\,
\begin{pmatrix}\Omega_k&0\\0&\Omega_{g-k}\end{pmatrix}\,\in\HH_g\,:\,
\Omega_k\in \HH_k,\;\Omega_{g-k}\in\HH_{g-k}\,\right\}\;\cong\;\HH_k\times\HH_{g-k}.
\ees
\end{enumerate}
The third constraint says that the restriction of the functions $\Xi_8^{(g)}[\Delta^{(g)}]$ to 'reducible' period matrices is a product of the corresponding functions in lower genus. These three constraints are quite the same of those of D'Hoker and Phong in \cite{DP6} for the functions $\Xi_6^{(3)}[\Delta^{(3)}]$. The main difference is in the second constraint: their one is more restrictive because they imposed, in analogy of the genus two case, that the measure should be the product of a theta constant at the fourth power times a suitable form of weight six. This form under the action of the symplectic group can take a sign depending on $M\in\modular(6,\ZZ)$ and on the characteristic $\Delta^{(3)}$ in the same way that the $\theta[\Delta^{(3)}]$'s do.  This implies that the factorized expression $\theta[\Delta]^4\Xi_6[\Delta]$ transforms as the $\Xi_8[\Delta]$: each term in the product transforms with a factor $\epsilon(M,\Delta)^4$, but $\epsilon(M,\Delta)^{4+4}=1$.
Conversely, if each $\Xi_8[\Delta]$ were a product of $\theta[\Delta]^4$ and another function, these other functions would satisfy constraint (2) of D'Hoker and Phong.

A fine analysis of the three constraints shows that they are quite redundant and they can be simplified and imposed on just one function, say $\Xi_8[0^{(g)}]$ where $0^{(g)}$ is the characteristic with all entries equal to zero, and from it, using the transformation constraint, one can define the other $2^{g-1}(2^g+1)-1$ functions. We do not report this reduction here, but the interested reader can find all the details in \cite{CDG}, section 2.5, 2.6 and 2.7.
In conclusion, the three constraints for the function $\Xi_8^{(g)}[0^{(g)}]$ are:
\begin{enumerate}
\item
the function $\Xi_8[0^{(g)}]$ is holomorphic on $\HH_g$;
\item
the function $\Xi_8[0^{(g)}]$
is a modular form of weight $8$ on $\Gamma_g(1,2)$;
\item
(1) for all $k$, $0< k <g$, and all $\tau_{k,g-k}\in \Delta_{k,g-k}$ we have
$$
\Xi_8[{}^0_0](\tau_{k,g-k})\,=\,
\Xi_8[{}^0_0](\tau_{k})\Xi_8[{}^0_0](\tau_{g-k});
$$
(2) if $\Delta^{(g)}=[{}^{ab\ldots}_{cd\ldots}]$ with $ac=1$ then
$\Xi_8[\Delta^{(g)}](\tau_{1,g-1})=0$.
\end{enumerate}
The group $\Gamma_g(1,2)$ is a finite subgroup of $\modular(2g,\ZZ)$ defined as the stabilizer of the null characteristic:
\begin{align*}
\Gamma_g(1,2):&=\{M\in \modular(2g,\ZZ):\; M\cdot[{}^0_0]\equiv[{}^0_0]\;\mbox{mod}\,2\}
\\
&=\{M\in \modular(2g,\ZZ)_g:\;{\rm diag}A{}^tB \equiv {\rm diag }C{}^tD\equiv \,0\;{\rm mod}\,2\,\}.
\end{align*}
The condition of the holomorphicity if $\Xi_8^{(g)}[0^{(g)}]$ is implicit in the definition of modular form, but we report it for clarity.

\section{Construction of the measure}
One can prove \cite{CDG,DG} that the three constraints imply that the form $\Xi_8[0^{(g)}]$ belongs to the vector space $V_\Gamma$ of the form that are left invariant by the action of the group\footnote{More precisely the group is the quotient $\Gamma_g(1,2)/\{M\in Sp(2g,\ZZ):\;A\equiv D\equiv I,\;B\equiv C\equiv 0\;\mbox{mod}\;2\,\}$.} $\Gamma_g(1,2)$. A function $f$ belongs to $V_\Gamma$ if it is a modular form with respect to a certain finite subgroup of $\modular(2g,\ZZ)$ and it is invariant for the action of $\Gamma_g(1,2)$, i.e. if $\rho(M)f=f$, where $M\in\Gamma_g(1,2)$ and $\rho$ is a representation of $\modular(2g,\ZZ)$ on the space of modular form, given by:
\be
(\rho(M^{-1})f)(\Omega)\,:=\,\det(C\Omega+D)^{-k}f(M\cdot \Omega).
\ee
Using the theory of representations of finite groups we computed \cite{DG} the dimension of the space $V_\Gamma$ for $g=1,2,3$ and, using a different approach Oura computed its dimension for $g=4$. In Table 2 we report these results.
\begin{table}[!h] \label{tab:dim}
\begin{center}
\bes
\begin{array}{ccccc}
\hline
g &  1 & 2 & 3 & 4 \\
\hline
\dim V_\Gamma & 3 & 4 & 5 & 7 \\
\hline
\end{array}
\ees
\end{center}
\caption{Dimensions of the space $V_\Gamma$ for $g=1,2,3,4$.}
\end{table}
Now, it is not hard to construct a basis for the space $V_\Gamma$ for $g=2,3,4$. The basis functions obviously satisfy the first two constraints, and only the third must be then imposed. The generic vector in $V_\Gamma$ is:
\be
\tilde{\Xi}_8^{(g)}[0]=\sum_i^n a_i^{(g)}e_i^{(g)} \qquad a_i\in\CC,
\ee
where $e_i^{(g)}$'s are the elements of the basis and $n=4,5,7$ for $g=2,3,4$ respectively. To determine the coefficient $a_i$ we impose the factorization constraint (the third):
\be
\Xi_8^{(g)}[0](\Omega_{k,g-k})=\Xi_8^{(k)}[0](\Omega_{k})\Xi_8^{(g-k)}[0](\Omega_{g-k}).
\ee
We can repeat this procedure iteratively on the subfactors $\Omega_g$ and $\Omega_{g-k}$, until we recover the well known expression for the amplitude at $g=1$.
The factorization can be computed observing that the theta constants factorize in a very simple way $\theta[{}^{a_1\cdots a_g}_{b_1\cdots b_g}](\Omega_{k,g-k})=\theta[{}^{a_1\cdots a_k}_{b_1\cdots b_k}](\Omega_{k})\theta[{}^{a_{k+1}\cdots a_g}_{b_{k+1}\cdots b_g}](\Omega_{g-k})$ and it is zero if $\Delta^{(k)}$ or $\Delta^{(g-k)}$ is odd.

We verified that basis for the spaces $V_\Gamma$ are:
\begin{itemize}
\item
g=2
\begin{align*}
F_1=\theta[0]^{16} && F_2=\theta[0]^4\sum_{\Delta} \,\theta[\Delta]^{12} && F_3=\theta[0]^8\sum_{\Delta}\theta[\Delta]^8 && F_{16}=\sum_{\Delta}\theta[\Delta]^{16}.
\end{align*}
\item
g=3
\begin{align*}
&F_1=\theta[0]^{16} && F_2=\theta[0]^4\sum_{\Delta} \,\theta[\Delta]^{12} && F_3=\theta[0]^8\sum_{\Delta}\theta[\Delta]^8 \\ 
&F_{16}=\sum_{\Delta}\theta[\Delta]^{16} && F_{88}=\sum_{(\Delta_i,\Delta_j)_o}\theta[\Delta_i]^8\theta[\Delta_j]^8.
\end{align*}
\item
g=4
\begin{align*}
&F_1=\theta[0]^{16} && F_2=\theta[0]^4\sum_{\Delta} \,\theta[\Delta]^{12} && F_3=\theta[0]^8\sum_{\Delta}\theta[\Delta]^8 \\ 
&F_{16}=\sum_{\Delta}\theta[\Delta]^{16} && F_{88}=\sum_{(\Delta_i,\Delta_j)_o}\theta[\Delta_i]^8\theta[\Delta_j]^8 \\
&F_{8}=(\sum_{\Delta}\theta[\Delta]^8)^2 && G_3[0] .
\end{align*}
\end{itemize}
The sums run over all the even theta characteristics for each genus. The sum of $F_{88}$ runs over all the couples of different non zero characteristics whose sum is odd (for the genus three case there are 280 such couples). The $G_3[0]$ is a function, constructed using the notion of isotropic spaces, see \cite{CDG,CDG2}, of the form:
\bes
G_3[0]=\theta[{}^{0000}_{0000}]^2\theta[{}^{0000}_{0001}]^2\theta[{}^{0000}_{0010}]^2\theta[{}^{0000}_{0011}]^2\theta[{}^{0000}_{0100}]^2\theta[{}^{0000}_{0101}]^2\theta[{}^{0000}_{0110}]^2\theta[{}^{0000}_{0111}]^2+\ldots
\ees
And $\ldots$ stands for other 2024 terms of this kind. Each term in the sum is composed by eight squared theta  constants whose characteristics satisfy, for each choice of $\Delta=\left[{}^{a_1\,a_2\,a_3\,a_4}_{b_1\,b_2\,b_3\,b_4} \right]$ and $\Delta'=[{}^{a'_1\,a'_2\,a'_3\,a'_4}_{b'_1\,b'_2\,b'_3\,b'_4}]$:
\be
a_1b'_1+a_2b'_2+a_3b'_3+a_4b'_4+a'_1b_1+a'_2b_2+a'_3b_3+a'_4b_4 =0 \bmod 2.
\ee
A set of characteristics of this kind is called an isotropic subspace and it has dimension three.

Imposing the factorization constraint we obtain a unique solution for the coefficient $a_i$.
Finally, the forms $\Xi_8^{(g)}[0^{(g)}]$ for $g\leq 4$ are:
\begin{itemize}
\item
$g=2$
\bes
\Xi_8[0]=\theta[0]^4\Xi_6[0]=-\frac{2}{3}F_1-\frac{1}{3}F_2+\frac{1}{2}F_3.
\ees
\item
$g=3$
\newline
\bes
\Xi_8[0]=\frac{1}{3}F_1+\frac{1}{3}F_2-\frac{1}{4}F_3-\frac{1}{64}F_8+\frac{1}{16}F_{88}.
\ees
\item
$g=4$
\newline
\bes
\Xi_8[0]=\frac{2}{3}F_1+\frac{4}{3}F_2-\frac{1}{2}F_3+\frac{34}{224}F_8-\frac{1}{4}F_{88}-\frac{10}{7}F_{16}-2G_3[0].
\ees
\end{itemize}

\section{Conclusion and open problems}
To consider the problem of finding the chiral superstring measures we have taken the stance that the vacuum to vacuum amplitude should split as conjectured by D'Hoker and Phong:
\bes
\mathcal{A}=\int_{\mathcal{M}_g}
(\det\Imm\Omega)^{-5}\sum_{\Delta,\Delta'}c_{\Delta,\Delta'}
{\rm d}\mu^{(g)}[\Delta](\Omega)\wedge
\overline{{\rm d}\mu^{(g)}[\Delta'](\Omega)}.
\ees
This splitting, as pointed out in section \ref{super}, is not strongly supported by some geometric results, as the Mumford theorem for the bosonic case. Moreover it is derived working in local coordinates and actually there is not a proof for it which considers the global proprieties of the (super)Riemann surfaces considered and its (super)moduli space.
Nevertheless, some remarkable facts seem to provide some evidence for the correctness of this assumption.

We are able to prove the uniqueness for the solution found by D'Hoker and Phong at $g=2$ and for our solution at $g=3,4$. The proof is based on the consideration of section \ref{theta} on the possibility of write every modular form in a substantially unique way as polynomial in theta constants for $g\leq 3$. For the genus four case we made also the assumption that the amplitude should be polynomial in the theta constants.

Supersymmetry imposes that the vaccum to vacuum amplitude should vanish:
\bes
\mathcal{A}=\int_{\mathcal{M}_g}
(\det\Imm\Omega)^{-5}\sum_{\Delta,\Delta'}c_{\Delta,\Delta'}
{\rm d}\mu^{(g)}[\Delta](\Omega)\wedge
\overline{{\rm d}\mu^{(g)}[\Delta'](\Omega)}=0.
\ees
We proved this for the measure found, actually it happens that 
$\sum_\Delta \Xi_8^{(3,4)}[\Delta](\Omega)=0$.
From this one deduces that $\sum_{\Delta}
{\rm d}\mu[\Delta](\Omega)=0$ 
and also that $\mathcal{A}=0$. Here the functions $\Xi_8^{(3,4)}[\Delta^{(3,4)}]$ are defined using the transformation constraint, for details see \cite{CDG}.

At $g=3$ it was proved \cite{GSM} that the 2-point function vanishes, as expected.

Behind these positive checks there are also some open problems when one tries to generalize this construction to higher genus.
A formal candidate for the forms $\Xi_8$ for $g\geq 5$ was proposed in \cite{G}  where some expressions appear but it is not clear if they are well defined for $g> 5$. Recently in \cite{OPSMY} a candidate for the genus five amplitude was proposed. There, the authors use a slightly different formalism: they use the theta series in place of the classical theta functions. It is not clear if these two formalism for $g\geq 5$ are equivalent and if one takes into account the same vector space. Moreover, in this case the vacuum to vacuum amplitude no more identically vanishes in a natural way. The vanishing of $\mathcal{A}$ must be imposed by hand. A recent discussion about the $g=5$ case is done in \cite{morlat}. This problem persists also for $g>5$.

The algebraic proprieties of the space of modular forms for $g\geq 5$ are not actually known, thus it cannot be proved the uniqueness of the $\Xi_8$ for $g\geq 5$. 

It is not clear if the two point function vanishes for $g>3$ and if the three point function vanishes for $g\geq 3$, as one expected from some general renormalization theorems.

From this considerations it is clear that the main issue is a mathematically rigorous proof of \eqref{1punto1}, which we accepted as correct. A deep investigation on the derivation of that expression would clarifies the outlined problems.

\section{Acknowledgements}
The author is grateful to S.\ L.\ Cacciatori for several stimulating discussions and for suggestions.

\end{document}